\def\year{2022}\relax
\documentclass[letterpaper]{article} % DO NOT CHANGE THIS
\usepackage{aaai21}  % DO NOT CHANGE THIS
\usepackage{times}  % DO NOT CHANGE THIS
\usepackage{helvet} % DO NOT CHANGE THIS
\usepackage{courier}  % DO NOT CHANGE THIS
\usepackage[hyphens]{url}  % DO NOT CHANGE THIS
\usepackage{graphicx} % DO NOT CHANGE THIS

\usepackage{makecell}

\usepackage{natbib}  % DO NOT CHANGE THIS AND DO NOT ADD ANY OPTIONS TO IT
\usepackage{caption} % DO NOT CHANGE THIS AND DO NOT ADD ANY OPTIONS TO IT
\title{Glowing Experience or Bad Trip? \\ A Quantitative Analysis of User Reported Drug Experiences on Erowid.org}
\author{
    Angelina Mooseder,\textsuperscript{\rm 1} Momin M. Malik,\textsuperscript{\rm 2} Hemank Lamba,\textsuperscript{\rm 3}\\
    Earth Erowid,\textsuperscript{\rm 4}
    Sylvia Thyssen,\textsuperscript{\rm 4}
    J\"{u}rgen Pfeffer\textsuperscript{\rm 1}    \\
}
\affiliations{
    \textsuperscript{\rm 1}Bavarian School of Public Policy, Technical University of Munich\\
    \textsuperscript{\rm 2}Berkman Klein Center for Internet \& Society at Harvard University\\
    \textsuperscript{\rm 3}School of Computer Science, Carnegie Mellon University\\
    \textsuperscript{\rm 4} Erowid.org, Grass Valley, California, United States\\
}
\newcolumntype{C}[1]{>{\centering\let\newline\\\arraybackslash\hspace{0pt}}m{#1}}

\begin{document}

\maketitle

\begin{abstract}
Erowid.org is a website dedicated to documenting information about psychoactive substances, with over 36,000 user-submitted drug Experience Reports. We study the potential of these reports to provide information about characteristic experiences with drugs. First, we assess different kinds of drug experiences, such as `addiction' or `bad trips'. We quantitatively analyze how such experiences are related to substances and user variables. Furthermore, we classify positive and negative experiences as well as reported addiction using information about the consumer, substance, context and location of the drug experience. While variables based only on objective characteristics yield poor predictive performance for subjective experiences, we find subjective user reports can help to identify new patterns and impact factors on drug experiences. In particular, we found a positive association between addiction experiences and dextromethorphan, a substance with largely unknown withdrawal effects. Our research can help to gain a deeper sociological understanding of drug consumption and to identify relationships which may have clinical relevance. Moreover, it can show how non-mainstream social media platforms can be utilized to study characteristics of human behavior and how this can be done in an ethical way in collaboration with the platform providers.
\end{abstract}

\section{Introduction}
In the past decade, social media research has primarily focused on large platforms like Twitter or Facebook to analyze social phenomena. This has lead to over-studied platforms in which challenges of generalizability are well-known \cite{Ruths2014}. In contrast, studying unique, small online communities can provide new insights into the breadth of human behavior. In particular, it can offer the opportunity to examine less publicly discussed aspects of human experience, such as drug\footnote{We refer to 'drug' as a term for the wide range of unapproved recreational, steroidal, performance enhancing, sedative, or other bio-active chemical technologies and the disapproved use of approved pharmacological agents of any variety.} consumption. 

Around the world, the consumption of drugs and the number of available substances has risen dramatically over the last decade and since 2005, around 950 new psychoactive substances have been identified worldwide \cite{world_drug_report_2020}. The great complexity of the world's drug landscape poses new challenges for social workers, governmental institutions, medical personnel, and drug consumers themselves \cite{arillotta_2020,dagnone_2015,schifano_2020}. It is important to detect adverse reactions to substances, understand under which circumstances they arise and implement meaningful harm reduction approaches, as well as to identify positive effects of specific drugs and understand their therapeutic potential. To achieve these goals, information is needed about why, when and in which context certain drugs are used and which effects they produce in different settings. 

One website providing such information is Erowid.org. For more than 25 years, the site has been collecting and curating information about (often illicit) drugs, serving people who use these substances, as well as family members, clinicians, educators, policy makers, and the curious general public. Of particular interest are the Experience Reports published on Erowid: in more than 36,000 reports, users\footnote{We refer to (Erowid) users as term for people, who submit Experience Reports to Erowid.} described what happened when they consumed one or more of over 800 different substances. Analyzing these reports gives a unique opportunity to study drug use from the consumer's perspective.

In cooperation with the Erowid staff members, we investigate the subjective experiences of drug consumers in a large-scale, quantitative manner. We first give a description of the dataset (Section 3) and the user base (Section 4). We then identify how various drugs and user variables are linked to characteristics of the drug experience (Section 5). Furthermore, we test whether the subjective outcome of a drug experience is predictable, using information about the drug, consumer and situational factors (Section 6). Based on a dataset of 36,711 user Experience Reports we find that:
\begin{itemize}
\setlength\itemsep{0.08 em}
\item { \textbf{Age plays a significant role in the motivation for and evaluation of drug consumption.} Younger people report more about bad and difficult experiences with drugs, while older people report more about using drugs for medical purposes.}
\item {\textbf{Males and females differ in their drug experiences.} Females report more about using drugs for medical purposes, while at the same time report more about health problems and addiction associated with drugs.}
\item {\textbf{The outcomes of reported drug experiences are linked to the substances consumed.} LSD is associated with negative experiences, while MDMA is associated with positive experiences.}
\item {\textbf{Reported health problems and addiction are linked to specific substances.}  The data revealed understudied patterns, such as an association between DXM and addiction, which may have clinical relevance.}
\item {\textbf{Although drugs and situational factors are correlated to the outcome of an experience, they do not yield enough information to predict whether an experience will be positive, negative, or associated with addiction.}}
\end{itemize}
We then discuss the benefits and limitations of using subjective drug Experience Reports found online. Our research presents new approaches for psychopharmacological research and can help to obtain a deeper sociological understanding of drug consumption. 

\section{Background and Related Work}

\subsection{Analysis of Drug Consumption}
The criminalization of substance abuse makes it difficult to accurately study drug consumption. Researchers often rely on surveys of focus groups \cite{nsdhu_2018,eu_drug_report_2020}, which suggests a high risk of reporting bias: respondents may not feel comfortable expressing their opinion about drug use and may conceal information about behavior which is illegal or not socially accepted. While collections of anonymous, self-reported experiences with drug usage are by no means a representative sample and in fact are likely a highly biased sample, we assume that their content is relatively free of self-reporting bias. Therefore, we use information substance consumers have voluntarily revealed in online communities to contribute new knowledge about drug consumption.

\subsection{Online Drug Communities}
Erowid.org was founded in 1995, early in Web history, but it is not the only drug-related Internet community. There are other communities that allow users to discuss substance related topics,\footnote{https://drugs-forum.com/}, exchange recipes and recommendations for substance use,\footnote{https://bluelight.org/} and publish information about the assumed contents of substances they have received.\footnote{https://pillreports.net/}
Many of these forums have the aim to offer accurate information about substances, share experiences of both positive and negative effects, give advice or warnings about dosages, and provide support for users experiencing negative reactions \cite{soussan_2014, rolando_2019}. By reading such information, users try to gain new knowledge about substances and to minimize their risk of experiencing adverse effects \cite{norman_2014, duxbury_2018,bilgrei_2019,berning_2016}.

The analysis of such communities has been used for detecting new trends on the drug market \cite{arillotta_2020, schifano_2020, rhumorbarbe_2019}, finding common drug-drug combinations \cite{chary_2018}, and understanding the use of specific substances \cite{andersson_slippery_2017, bonson_1996}. 

\subsection{Erowid.org}
The website Erowid.org has become a valuable resource for researchers seeking to gain information about the compounds, dosages, classification and effects of drugs, especially for substances with no or little medical documentation \cite{karila_2016, stanciu2016}. The organization Erowid Center has used the website to recruit participants for surveys on visitor demographics and experiences with certain substances\footnote{https://www.erowid.org/general/survey/} as well as for surveys on specific drugs conducted in collaboration with academic researchers \cite{pal_2013,baggott_2011,baggott_2010,gamma_2005}. 

The experience reports published on Erowid have been qualitatively analyzed to gain an understanding of the use of synthetic cannabinoids and Kratom \cite{swogger_2015, newman_2016}. Furthermore, these reports have been used for anthropological case studies, in which the authors visualized patterns, such as substance co-use, common dosages of specific drugs or wordpair-substance relationships \cite{krieg_2016, krieg_2017}. However, there still exist many more use cases on which such data could be applied. The aim of this paper is to understand the potential of subjective drug consumption reports to provide information about the characteristic experiences of these drugs. 

\subsection{Ethical Considerations}
We are conducting research on open (though anonymous or pseudonymous) admissions of sometimes illegal behavior, and we have a responsibility to not put together data in such a way that it would deanonymize individuals and potentially put them at risk. 
When submitting reports, users explicitly grant permission to use their reports for scientific research. Therefore, care is already taken in the writing and editing process to omit or obscure identifiable details. 
Our data contains no personal identifiable information (PII). Nothing in our models tries to fill in or infer information from individual reports that could help identify the authors. 

Second, the manufacture, distribution, possession and/or consumption of many of the substances described in Erowid Experience Reports are illegal in many or most jurisdictions. By studying these reports in a nonjudgmental way, we risk abetting and condoning behavior that could be illegal and that many consider immoral. However, there is a long precedent in sociology and anthropology of studying such behavior; for drugs specifically, we work within the educational and \textit{harm reduction} frameworks \cite{marlatt1996} supported by Erowid and by many biomedical researchers, social workers, and medical practitioners. 
Harm reduction is an evidence-based approach built on more than three decades of empirical research  \cite{stone2018}, and it prioritizes providing accurate, judgement-free information, safe environments for drug consumption, addiction treatment, and decriminalization of drug usage and markets, all of which lead to less harm from drug consumption.
The current study obtained the explicit consent of Erowid Center before downloading reports off their website. Furthermore, we recognize the labor and expertise of the Erowid staff members through co-authorship.

\section{Data}
\subsection{Erowid}
Erowid is viewed by advocates of drug policy reform and harm reduction as one of the most important resources on drugs \cite{jarnow_2016}. It describes itself as
\begin{quote}``a member-supported organization providing access to reliable, non-judgmental information about psychoactive plants, chemicals, and related issues. We work with academic, medical, and experiential experts to develop and publish new resources, as well as to improve and increase access to already existing resources. We also strive to ensure that these resources are maintained and preserved as a historical record for the future.'' \footnote{https://www.erowid.org/general/about/about.shtml} \end{quote}  

Much of the site is devoted to psychoactive substance `vaults', subsections containing extensive information about various psychoactive drugs. The `experience vaults',\footnote{\label{experience_vault}https://erowid.org/experiences/} a collection of narratives about consuming psychoactive substances (and/or practicing psychoactive methods, such as fasting or meditation), largely submitted by site readers (with additional Experience Reports republished and compiled from other sources), is particularly fascinating.

Over a hundred thousand experience reports have been submitted to Erowid, about 35\% of which have been published. Reports pass through a review process, with each report being reviewed by two out of a few dozen trained volunteers who read and rate the submissions and pass them on to senior reviewers for a possible publication \cite{erowid_30,witt_2015,erowid_10}. During the review process, each report is tagged with `primary' categories (= type of experience, such as `Bad Trip') and `secondary' categories (= situational factors, such as `Nature/Outdoors') \cite{erowid_30}. The standardized format of Experience Reports, with fields where users input substance name, dosage, and form of consumption (pill, smoked, etc.), as well as the quality control through Erowid reviewers make them particularly valuable compared to similar sites.

\subsection{Data Collection}
Following the rules of spidering given on the Erowid site,\footnote{https://erowid.org/general/about/about\_archives1.shtml} we contacted Erowid and received permission to crawl the site. We then collected 36,778 html pages between 2021-02-16 and 2021-04-07. As 65 of these pages did not follow the formatting standards of Experience Reports, they were omitted, leaving a dataset of 36,713 reports. Each report generally consisted of an identification number, title, text, publication date, author information, such as pseudonym, age,\footnote{The age field was included in 2009.} gender and weight, substance information, dosage information, number of views, year of experience, and category labels assigned by the Erowid team. 

\subsection{Data Cleaning}
\label{section:data_clean}
\subsubsection{User and Report Information.}

We cleaned and regularized the data extensively. We manually inspected all reports with an unusual author age under 13 or over 70. In cases where the report showed indications that the age information was wrong (eg. the age being four-digit, or the activities not matching the age, such as a 12-year-old being the driver of a car), we replaced the age with a missing value.
We converted the user weight information, which was given in pounds (lb), kilograms (kg), or stones (st), into pounds. We replaced unrealistic low (\textless 70 lb) and high (\textgreater 500 lb) weights with a missing value. The inspection of reports with unrealistic weights revealed two reports about a cat and a dog, which had unintentionally consumed drugs. These reports were excluded from the dataset, resulting in a set of 36,711 reports. Reports on Erowid were not assigned with a unique user ID, but only with a pseudonym. Each user could have multiple pseudonyms and one pseudonym could be assigned to multiple users. Hence, it is not possible to evaluate the exact number of users and their submission history. When assessing user demographics, we counted each report as one user to obtain an upper estimation of demographics.

\subsubsection{Categories and Context.}
To assess the characteristic experience of a report, we stored all `primary' categories, which were assigned to a report by Erowid reviewers. 
To assess the context and location of a drug experience, we used `secondary' category labels and context labels which were also assigned to reports by Erowid reviewers: We created the feature `Party', which is 1 if a report contains the label `Large Party', `Club/Bar', `Rave/Dance Event' or `Festival/Large Crowd'. We constructed the feature `Therapeutic', which is 1 if the report contains the label `Therapeutic Intent or Outcome' or `Therapeutic Session'. We furthermore added binary features based on the labels `Workplace', `School', `Public Space', `Nature/Outdoors', `Guides/Sitters', `Alone' and `Multi-Day Experience'.

\subsubsection{Substance Information.}
For each report, we used the dosage field as the source of substance information, as this provided additional information about dosage amount and consumption method. The distribution of the top 100 drugs in our dataset can be found in section A of an appendix we published online.\footnote{\label{foot:appendix}https://mediatum.ub.tum.de/1639245} The dataset comprises reports of about 845 distinct substances (674 when grouped into larger classes).  However, only 44\% of these substances were included in more than ten reports.  We decided for all analyses, where the drug was an independent variable (see Section 4.2, 5.2, 6), to focus only on reports about the ten most popular drugs. The ten most reported substances were `Cannabis', `MDMA', `Salvia Divinorum', `LSD', `Mushrooms', `Alcohol -- Beer/Wine', `Tobacco -- Cigarettes', `DMT', `DXM' and `Ketamine'. As `Mushrooms' is a superclass for a great variety of psychedelic mushrooms, we decided to focus instead on the more specific class `Mushrooms - p. cubensis', which was the 15th most reported substance. Since both `Alcohol - Beer Wine' and `Tobacco - Cigarettes' were used in more than 90\% of cases in combination with other drugs, we decided not to include them among examined categories. Instead, we considered the 11th and 12th most reported substances, `cocaine', and `amphetamines'. As amphetamines is again a superclass, we included the most common amphetamine `methamphetamine' (ranked 20 of the most reported substances).
We then stored for each report whether it would contain one or more of our top ten drugs: We created one indicator feature for each drug (e.g. `mdma'). Each feature was assigned  1 if the respective report contained the drug and 0 otherwise. Furthermore, we constructed one indicator feature for each combination of drugs (e.g. `mdma-dxm' or `mdma-dxm-dmt'). As there were 10 drugs, we had 1024 possible combinations, although many combinations did not appear in the data set. From including only combinations with at least one report, our dataset yielded 142 drug and drug combination features. 

Table \ref{table:topten} shows the resulting single substances, the number of reports per substance and the percentage of all reports in the dataset. In our data, 15,861 reports include at least one of these top ten drugs, which accounts for 43\% of our dataset.

\begin{table}%[t]
\footnotesize 
\centering
%\resizebox{.95\columnwidth}{!}{
\begin{tabular}{ll|ll}
    Drug &  Reports & Drug &  Reports \\
    \hline
    Cannabis & 7,274 (20\%) & DXM & 843  (2\%) \\
    MDMA & 2,406  (7\%) & Ketamine & 780  (2\%) \\
    Salvia Div. & 2,319  (6\%) & Cocaine & 776  (2\%) \\
    LSD & 2,263  (6\%) & P. cubensis & 657  (2\%) \\
    DMT & 943  (3\%) &  Meth. & 553  (2\%)
\end{tabular}
\caption{Top Ten Drugs used for the current study with number of reports and their percentage in the dataset. Reports describing the use of multiple substances were counted once for each drug.}
\label{table:topten}
\end{table}

\subsubsection{Dosage Information.}
After extracting the substance information for each report, we were also interested in the dosage amount. However, extracting dosages proved to be a challenging task for several reasons. First, inconsistent and incommensurable measures were used: grams and ounces, but also `tablets', `bowls', `shots', `lines', `cookies', `drops' and more. Since the same volume or weight can have different concentrations of an active substance, even with domain expertise the standardization of these dosage amounts would not be possible. Second, the dosage was often described in approximations such as `multiple' or `some' [unit]. Third, there were numerous methods to consume a certain drug (e.g. smoking, drinking,...) and it was not always possible to generate rules about how a dosage with one consumption method can be converted in a dosage with another consumption method. Fourth, there were numerous forms of substances per drug with varying levels of the active constituent and therefore the dosage amount had to be adapted to the substance form. Finally, there was a huge amount of missing data for dosages. For example, more than 60\% of cocaine reports did not include dosage information. 
To gain at least a rough estimation of dosages, we selected for each of our top ten drugs a reference consumption method, unit, and substance form. Using the information about common doses given on various websites,\footnote{https://www.erowid.org/, https://www.trippingly.net/, \\
\phantom{}\ \ \ \ \ \ \ https://drugs.tripsit.me, https://dancesafe.org} we created a set of heuristics per drug about how to convert dosages of other units, consumption methods or substance forms. For smoked drugs, we generally converted a `bowl' to the height of one common dose and estimated a `hit' to be one third of a bowl. All data, for which we could not infer any rules about either unit, consumption method or substance form, was set to a missing value and ignored for all dosage analyses. The heuristics and an overview of the most common terminologies used can be found in section B in the online appendix.\footnotemark[11] Table \ref{table:dosagedis} shows for each drug the percentage of reports where dosage information was given and the distribution of dosages in these reports. It should be noted that the maximum dosage values can be quite high, as some users reported the amount of drugs they had taken over several days or together with other consumers.

\begin{table}
\footnotesize
\centering
%\resizebox{.95\columnwidth}{!}{
\begin{tabular}{l|l|l}
    Drug &Perc& Dosage distribution\\
    \hline
    Cannabis & 35\% & 15.0 - 142k (M=486, SD=3,029)\\
    MDMA & 76\% & 0.50 - 160k (M=284, SD=3,810)\\
    Sal. Div. & 75\% & 0.03 - 500k (M=1,932, SD=12,246)\\
    LSD & 87\% & 10.0 - 5k (M=250, SD=279)\\
    DMT & 68\% & 0.08 - 1k (M=51, SD=67)\\
    DXM & 80\% & 0.60 - 510k (M=18,971, SD=67,538)\\
    Ketamine & 45\% & 0.16 - 260k (M=986, SD=13,929)\\
    Cocaine & 14\% & 0.12 - 11k (M=1,350, SD=1,568)\\
    P. cub. & 81\% & 1.75 - 170k (M=4,015, SD=7,707)\\
    Meth. & 20\% & 5.00 - 25k (M=668, SD=2,719)\\
\end{tabular}
\caption{Top Ten Drugs used for the current study with percentage of reports where dosage information was given and distribution of dosage information. LSD is reported in ug, all other drugs in mg.}
\label{table:dosagedis}
\end{table}

\subsection{Dataset Overview}
Table \ref{table:summary} gives an overview of the dataset, regarding the number of reports and substances, distribution of number of substances used per report, as well as distribution of report views, drug experience year, user age, weight and gender.

\begin{table}[h]
\footnotesize
\centering
\begin{tabular}{l|l}
    Variable &  Statistics \\
    \hline
    Reports & 36,711 \\ 
    Substances & 845 \\
    Drugs p. r. & 1-13 (M=1.62, SD=1, P=100\%) \\
    Views & 74-777k (M=15,711, SD = 25k, P=100\%)\\
    Year & 1848 - 2021 (M=2,007, SD=6, P=99\%) \\
    Age & 7-80 (M = 25, SD=9, P=33\%) \\
    Weight &  70-500 (M=162, SD=37, P=93\%)\\
    Gender & Male=29,052, Fem.=5,449, Not Spec.= 2,210\\
\end{tabular}
\caption{Dataset Overview. P stands for the percentage of reports containing that information.}
\label{table:summary}
\end{table}

\section{Trends on Erowid}
\label{section: Trends on Erowid}
In this section, we analyze the whole dataset of 36,711 reports to identify 1) which drug trends are prevalent on Erowid and 2) what kind of users report to Erowid. Following \citeauthor{paul_2016}'s (\citeyear{paul_2016}) work on drugs-forum.com, we compare these trends to national and international estimations of drug usage to obtain a basic understanding of the data source and the group of Erowid contributors.

\subsection{Drug Popularity}
Table \ref{table:topten} shows the 10 most common substances (as defined in section 3.3) and their distribution in the dataset. 

First, we find that \textit{Erowid contributors show a higher interest in substances outside the most commonly used drugs}. On a global level, the estimated distribution of substance use is quite skewed towards cannabis: According to the UN's estimations, 71\% of past-year ‘drug’ users, have consumed cannabis, which makes it by far the most consumed substance worldwide, excluding alcohol and tobacco. Other commonly used substances are MDMA/ecstasy (21\%), amphetamine and methamphetamine (10\%) as well as cocaine (7\%) \cite{world_drug_report_2020}.
We find that the most common drugs are also popular on Erowid, although the distribution of drugs seems to be more balanced in our data: Comprising 20\% of reports, cannabis is the most prevalent drug on Erowid and also MDMA (7\%), methamphetamine (2\%) and cocaine show a comparatively high prevalence (2\%). The fact that these percentages are not higher indicates that the Erowid users also show a high interest in other substances, which are likely less prevalent on a global level.

Second, we find that \textit{this interest is especially strong regarding psychedelic substances}. Seven of the top ten reported drugs, namely MDMA, \textit{Salvia divinorum}, LSD, DMT, DXM, ketamine and \textit{Psilocybe cubensis} are substances which are categorized as hallucinogens by the North American National Survey on Drug Use and Health \cite{nsdhu_codebook_2018}. While in the USA the prevalence of hallucinogens is quite low (1\% of all past month substance use) in comparison to alcohol (85\%), marijuana (17\%) and cocaine (1\%) \cite{nsdhu_2018}, this is not the case among Erowid Experience Reports: 25\% of reports contain at least one of these seven hallucinogens. Therefore, the group of Erowid contributors likely does not correspond to global or even North American drug consumption but rather to consumption by a population with a strong interest in psychedelic experiences.

\subsection{Demographic Trends}
\begin{figure}[ht]
   \centering
   \includegraphics[width=1.0\columnwidth]{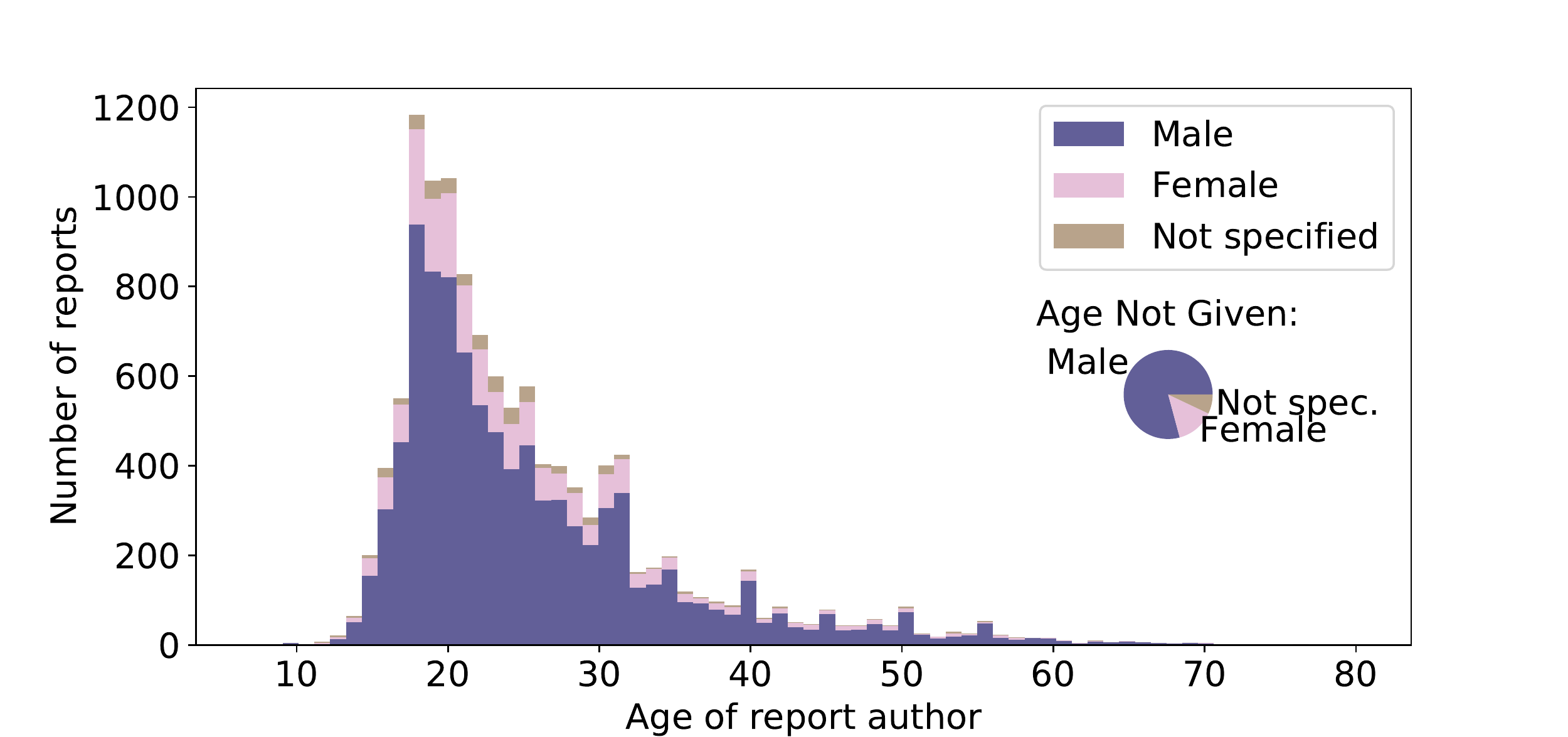}
   \caption{Stacked bar chart representing the age and gender distribution over all reports.}
   \label{fig:trend_agesex}
\end{figure}

Figure \ref{fig:trend_agesex} shows the age and gender distribution over all reports. Note that `gender' is coded only as `male', `female', or `not specified'; if users select the option "non-binary/other",  "prefer not to answer" or  remain with the default option ("- - -") when asked for their gender, this will be displayed as ‘not specified’ in the downloaded dataset.

First, we find that \textit{users on Erowid seem to have very similar age trends as estimated for global drug consumption}.
Surveys conducted worldwide show that drug consumption is more popular among younger people, with peak levels between 18 and 25 years \cite{world_drug_report_2018}. This trend is also prevalent in our data. Among the third of reports that contained a valid self-report of age, at the time of experience users were on average 25 years old, with 54\% of users being between 18 and 25 years old (min$=7$, max$=80$, MD$=22$, IQA$=19-29$). The distribution rises between ages 17 and 18, as the number of reports from 18 years olds is twice as high as the number from 17 years olds. Possible explanations are consumption differences, as drugs may be less accessible or interesting for minors, reporting bias, as users may feel more confident to report their age when over 18, reviewer bias, as some reviewers are less inclined to publish reports written by minors or quality-of-report bias, as writing quality or data content may be lower with juvenile authors, making reports by these authors less likely to be published.

Second, we find that \textit{the group of Erowid contributors is highly skewed towards males}.
Both in North America\footnote{Data from https://www.datafiles.samhsa.gov/} and the European Union \cite{eu_drug_report_2020}, drug consumption is reported more frequently by men than women, with around 60\% of those reporting drug use in the EU being male and around 40\% being female. The `gender' imbalance on Erowid is even greater: of all reports where gender is specified, 84\% are reported as `male'. Potentially, males are more interested in or willing to write about their own drug experiences or find the website more interesting than females.

\section{Characteristic Drug Experiences}
In this section, we 1) analyze which characteristics of a drug experience are described by categories given on Erowid and 2) identify associations between these characteristics and individual drugs as well as user variables. 

\subsection{Category Overview}
\label{section:cat_des}
Here we analyze which characteristics of a drug experience are expressed within a certain category. We first give a short description of each category and their distribution over all 36,711 reports. We then analyze the sentiments expressed in each category as well as category co-occurrence.

\subsubsection{Description.}
\begin{figure}[ht]
   \centering
   \includegraphics[width=1.0\columnwidth]{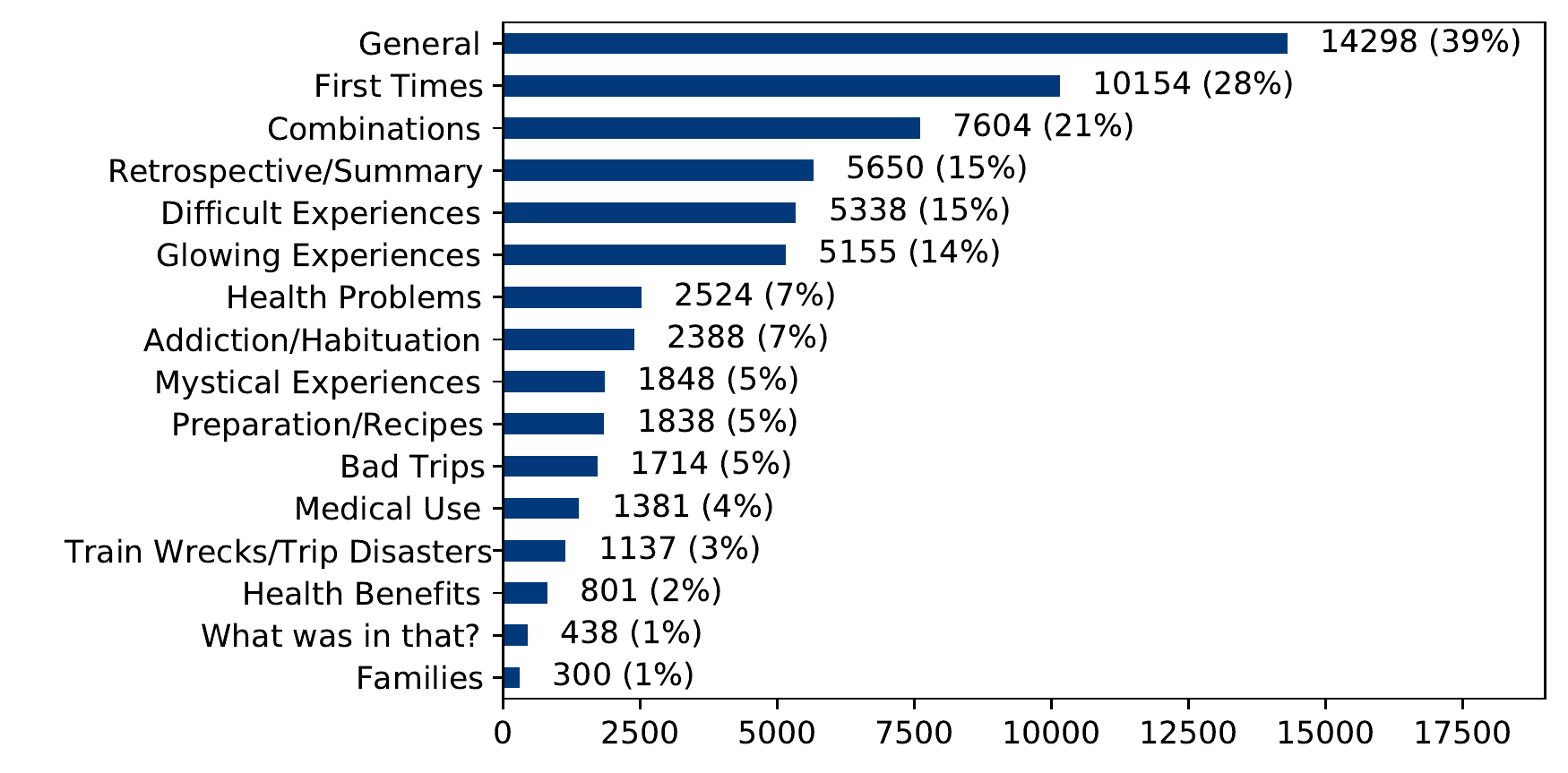}
   \caption{Number and percentage of reports per category in the dataset.}
   \label{fig:category_dis}
\end{figure}

During the review process conducted by Erowid, each report is assigned by the Erowid team one or more of 15 primary category labels. 
Figure \ref{fig:category_dis} presents the distribution of each category in the dataset.

The labels \textit{`Bad Trips'}, \textit{`Train Wrecks/Trip Disasters'} and \textit{`Difficult Experiences'} were assigned to reports about complicated, not entirely positive experiences during drug consumption, often caused by pharmacological reactions. %and leading to interference with the police or hospitals. 
In contrast, the label \textit{`Glowing Experiences'} was assigned to joyful experiences with drugs and the category \textit{`Mystical Experiences'} was dedicated to reports about psychoactive substance induced transcendent encounters.

The label \textit{`Medical Use'} was attached to reports on the consumption of substances for medical reasons, while the label \textit{`Health Problems'} was assigned to reports about medical issues in general. In both sections, users often reported adverse drug effects. In contrast, the \textit{`Health Benefits'} label was assigned to user experiences in which a substance helped to overcome certain health issues. The label \textit{`Addiction/Habituation'} was assigned to reports about drug dependence, and many of these reports were not about a specific event, but rather a longer period of time. The label \textit{`Retrospective/Summary'} was assigned to reports written in hindsight or over a longer period of time. The category \textit{`What was in that?'} was used for reports in which users suspected a discrepancy between the ingredients they thought their drug would include and the real composition of the drug. 

\textit{`General'} was a cumulative category and had no specific meaning. The label \textit{`First Times'} was assigned to reports about a person's first experience with a substance, and the label \textit{`Combinations'} was assigned when the consumption of more than one substance was the main topic of the report. The category \textit{`Preparation/Recipes'} was assigned to reports with a strong focus on the form of consumption. When reading the Experience Reports, it became evident that users on Erowid often reveal a high curiosity and experiment with new ways of consuming substances. The category \textit{`Families'} was assigned to a great variety of reports, with the common factor of family members being involved. This included reports about persons consuming drugs in the presence of or together with family members, about users thinking of family during the drug experience, about facing drug addiction with the help of family members, and more.

\subsubsection{Sentiments Expressed per Category.}
\begin{figure}[ht]
   \centering
   \includegraphics[width=1.0\columnwidth]{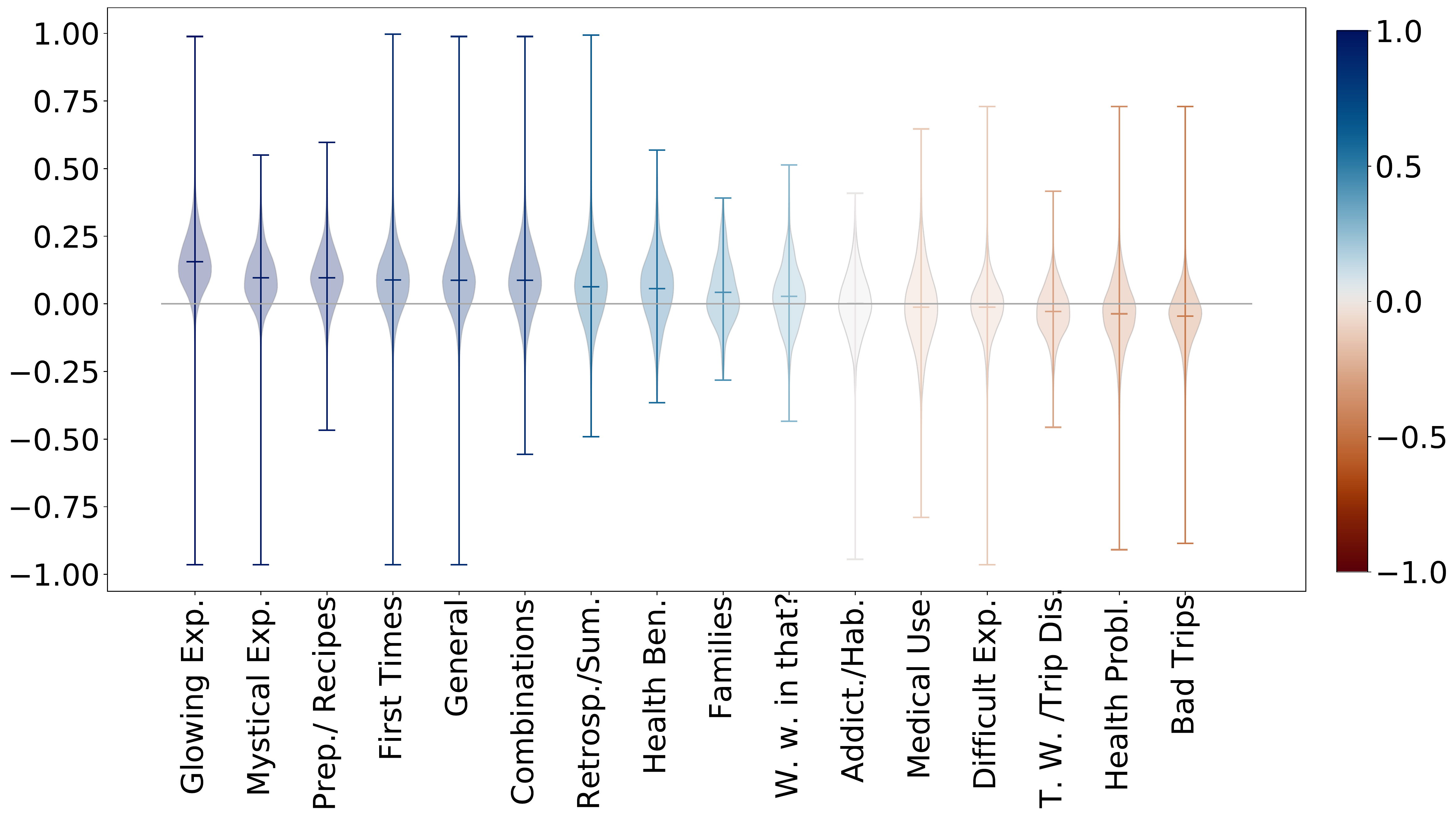}
   \caption{Distribution of sentiments per category. Lines mark the minimum, mean and maximum value. Color corresponds to the average sentiment score.}
   \label{fig:category_sents}
\end{figure}

To assess the sentiments described in a category, we used VADER \cite{Revision_Hutto2014}, a Python package which allows the calculation of sentiments expressed in a text and even includes the presence of negations, degree modifiers and more. We excluded the words `like', `ecstasy' and `funny' from the Vader dictionary, as they had different connotations in a drug specific context. We then lower cased the texts, calculated the Vader compound sentiment score (-1=fully negative, 1=fully positive) for each sentence in a report and stored the average compound score per report. Figure \ref{fig:category_sents} illustrates the distribution of sentiment scores per category. 

First, we find that \textit{labels assigned by the Erowid team correspond to the sentiments expressed in the categories}. As expected, reports in categories about negative drug experiences, such as `Bad Trips', `Train Wrecks/Trip Disasters' and `Difficult Experiences', show on average rather negative sentiments, while reports in the category `Glowing experiences' show on average rather positive sentiments. Moreover, the category `Health Problems' has a rather negative average sentiment score, while `Health Benefits' has a slightly positive average sentiment score. These results can be seen as a validation of Erowid's labelling process.

Second, we find that \textit{much positivity is expressed in categories about new and mystical drug experiences}. The categories `Mystical Experiences', `Preparation/Recipes', `Combinations' and `First Times' show on average positive sentiments. Research on similar platforms suggests that online community members sometimes show characteristics of so called `psychonauts': Their drug consumption is mainly motivated by curiosity about new substances and their possible applications, as well as the goal to gain knowledge about the inner self and the mysteries of life \cite{rolando_2019}. The fact that reports in these four categories show a high amount of positive sentiments supports the hypothesis that Erowid contributors are also often interested in learning about new substances, combinations and their preparation, and value the knowledge gained by spiritual experiences.

\subsubsection{Category Co-ocurrence.}

\begin{figure}[ht]
   \centering
   \includegraphics[width=0.9\columnwidth]{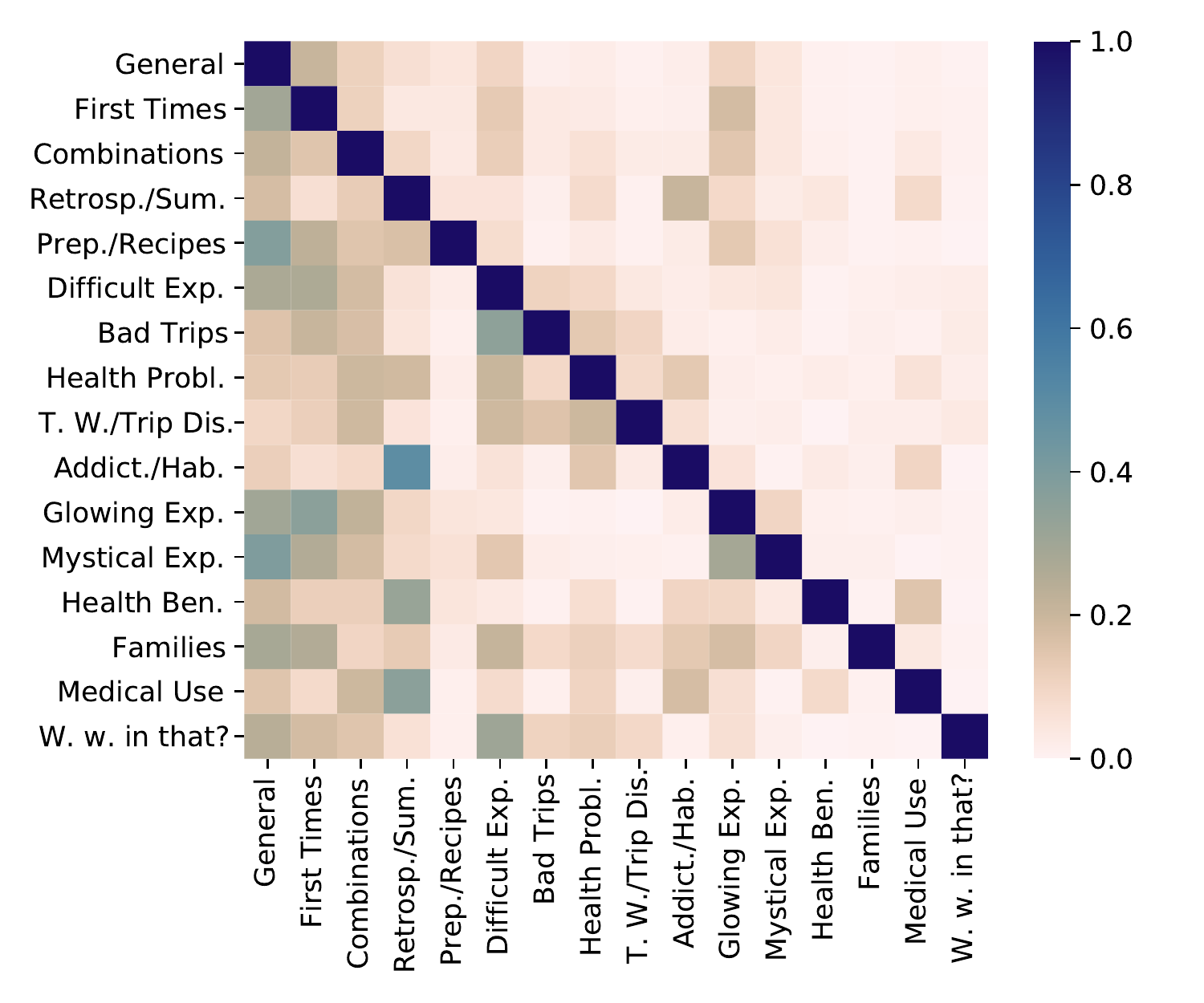}
   \caption{Category co-occurrences. The color in a cell $(i,j)$ represents the proportion of reports in category $j$, which are also assigned with category $i$ (i.e., co-occurrence frequencies are row-normalized). }
   \label{fig:category_coocurrence}
\end{figure}

To analyze category similarity we calculated the label co-occurrence in all reports, which is shown in Fig \ref{fig:category_coocurrence}. First, we find that \textit{categories, which we would expect to have a high topical overlap, show a high co-occurrence}. For example, there is a considerable co-occurrence between the categories `Difficult Experiences', `Bad Trips', `Health Problems' and `Train Wrecks/Trip Disasters'. Moreover, there is a considerable overlap between `Addiction/Habituation', `Health Benefits' and `Medical Use' with the category `Retrospective/Summary'. A qualitative inspection of these reports showed that many reports in the first three categories were not written about a specific event, but rather a longer period of time. Furthermore, the genre `What was in that?' often occurs in combination with `Difficult Experiences', which is not surprising, as unexpected drug effects may indeed lead to troublesome drug experiences. 

Second, we find that \textit{new and mystical experiences co-occur with joyful experiences}. There is a rather high overlap between `Mystical' and `Glowing' experiences. In line with the results of the sentiment scores, this suggests that Erowid contributors interpret transcendent encounters often as positive, valuable and/or joyful experiences. Moreover, we find a rather high co-occurrence between `First Times' and `Glowing Experiences'. Possible reasons for this are that Erowid users may find it joyful to experiment with new substances (corresponding to the description of psychonauts), that there may be a selection effect in that people who have positive initial experiences are more likely to submit reports about it to Erowid, and/or that there may be a causal relationship in that first experiences indeed are more frequently positive. 

In summary, we have shown that categories on Erowid describe a great variety of characteristic experiences. Categories range from positive drug experiences (e.g. `Glowing Experiences') to negative experiences (e.g. `Bad Trips') and even include special topics (e.g. `Addiction/Habituation'). These results provide us with a better understanding of categories and help to interpret associations between categories and drugs or user variables.

\subsection{Drug-Category Associations}
In this section, we analyze the associations between categories and the top ten substances in our dataset to determine whether characteristic drug experiences are linked to specific substances.

\subsubsection{Methods.} 
To examine the correlation between drug and category, we measured the chi-square distance, which is the difference between the observed cell frequency and the expected cell frequency under an independence hypothesis (the product of the row and column marginals), normalized by expected cell frequency. We limited this analysis to only the 15,861 reports which contained at least one of the top ten substances.

\subsubsection{Results.} 
\begin{figure}[ht]
   \centering
   \includegraphics[width=1.0\columnwidth]{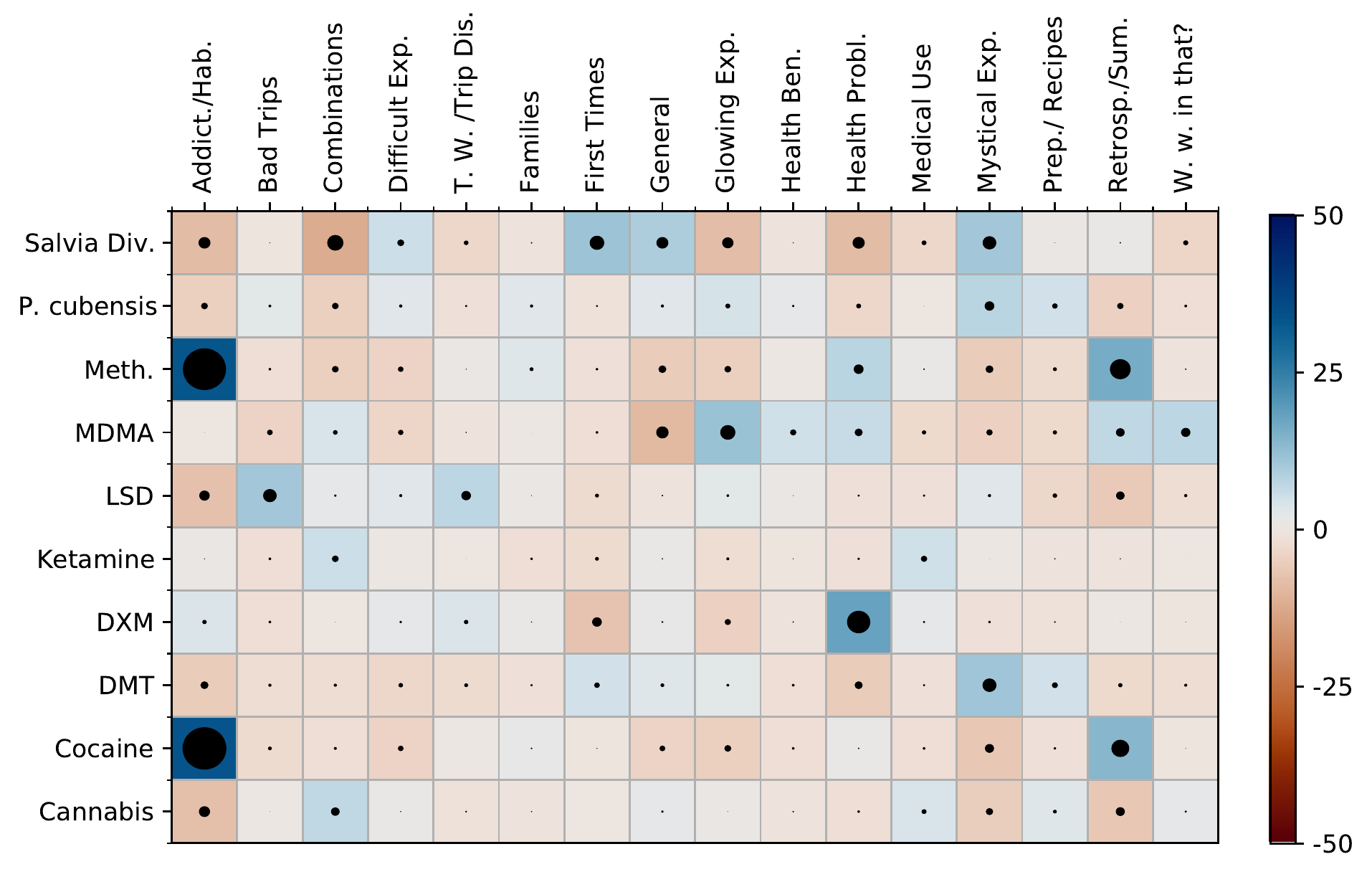}
   \caption{Chi-square distances of drug-category pairs. Positive cell values (blue) represent a positive correlation and negative cell values (red) a negative correlation between the given drug and the given category. The colors and circle sizes are proportional to the chi-squared distance; for example, the Addiction/Meth. pair has a distance of 33, Combinations/Salvia Div. of -12 and Bad Trips/Cocaine of -3.
   }
   \label{fig:cat_drug_residuals}
\end{figure}

There was a significant association between drug and category ($\chi^2(135, N=32,796)=5,924, p< 0.01$). Figure \ref{fig:cat_drug_residuals} shows the chi-square distance for each drug-category pair. The exact chi-square distances and contributions can be found in section C of the online appendix.\footnotemark[11]%\footnotemark[\ref{foot:appendix}]

First, we find that the \textit{subjective outcomes of drug experiences are linked to the substance consumed}. Key Findings (=findings with very high or low chi-square distances in comparison to values in the same row and column)  include:
\begin{itemize}
\item LSD correlates with negative experiences (`Bad Trips', `Train Wrecks/Trip Disasters').
\item MDMA correlates with positive experiences (`Glowing Experiences'). 
\item DMT, \textit{Salvia divinorum} and \textit{Psilocybe cubensis} correlate with `Mystical Experiences'.
\end{itemize}
The type of experience may be induced by the  pharmacological effects of the drug. For example, LSD often produces effects of paranoia and anxiety, MDMA often leads to euphoria, and DMT, \textit{Salvia divinorum} and \textit{Psilocybe cubensis} are known to produce hallucinogenic experiences. However, it is surprising that such correlations can be found in the data, as all of these substances also produce other pharmacological effects, which could lead to both positive and negative experiences. Furthermore, the effects of a psychoactive substance may also be influenced by the dosage, consumer variables, and the setting. Therefore we suggest that not only the biological effects play a role, but also the drug consumers expectations of these effects. For example, when users read that MDMA leads to `Glowing Experiences', they may expect to have a positive experience with MDMA, and put themselves in a position to enjoy such a positive experience (e.g. by meeting friends, dancing,..). This could then in turn positively influence the possibility of having a positive experience and interpreting it as positive.

Second, we find that \textit{reported health problems and addiction are linked to substances}. Key Findings include:
\begin{itemize}
\item Methamphetamine and cocaine correlate with `Addiction/Habituation'.
\item \textit{Salvia divinorum}, cannabis and LSD negatively correlate with `Addiction/Habituation'.
\item DXM correlates with `Health Problems' and weakly with `Addiction/Habituation'. 
\end{itemize}
The first two relations are in line with existing research, as both methamphetamine and cocaine are known to have a high potential for addiction, while for \textit{Salvia divinorum}, cannabis and LSD only weak to no withdrawal symptoms are known. Therefore, the Erowid data might help to compare drugs regarding their addiction potential or even detect worrisome relationships between certain substances and medical issues. For example, our data suggests that DXM may be related to health problems and addiction, which should be investigated through further research (see Section \ref{section:impl}). Such an approach could be especially beneficial for new substances, with not much clinical data available. 

Third, we find that \textit{specific usage patterns are linked to substances}. Key Findings include:
\begin{itemize}
\item Methamphetamine and cocaine correlate with `Retrospective/Summary'.
\item MDMA correlates with `What was in that?'.
\end{itemize}
The first relation suggests that users write more often about methamphetamine and cocaine in retrospect, than they do with other drugs. This can be explained by the high addiction potential of these drugs and the tendency to use these substances repeatedly over a period of time, which may lead consumers to write about this period instead of a specific recent event. In addition, there are certainly reasons for the second relation: MDMA is the active substance in most ecstasy tablets, which vary in content and concentrations. Consequently, users often under- or overestimate the MDMA concentrations or fail to detect dangerous additional substances \cite{vrolijk_2017}. While this particular connection is already known to users and medical professionals, it illustrates a real-world trend reflected in aggregate patterns within Erowid Experience Reports. 

In summary, our data reveals that the subjective outcomes of drug experiences, reported health problems and drug usage patterns are linked to specific substances. From a sociological perspective, this can help us better understand the motivations, expectations and usage patterns that drug consumers have with certain substances. From a medical perspective, these results reveal understudied patterns, such as the association between DXM and addiction, which may have clinical importance. 

\subsection{Associations between Category and User Variables}
In this section, we analyze the associations between categories and user variables to find out whether characteristic drug experiences are linked to consumers' age and gender.

\subsubsection{Methods.}
To examine the relation between age and category, we selected all reports with age information given ($n=11,993$), assigned an age group to each report (`Under 18', `18-25', `26-40', `Over 40') and performed a chi-square test of independence. Furthermore, again using chi-square distance, we analyzed the relationship between gender and category on all reports where gender information was specified ($n=34,501$). 

As there is a bias towards young, male users among Erowid contributors, the report numbers per group differ (`Under 18': 1,245, `18-25': 6,488, `26-40': 3,374, `Over 40': 886, Male: 2,9052, Female: 5,449). However, as the chi-square test compares the observed cell frequencies with the expected cell frequencies and the expected cell frequency is not below 5 in more than 20\% of cases, the differences in group size do not limit the significance of our results. 

\subsubsection{Results.}
\begin{figure}[ht]
   \centering
   \includegraphics[width=1.0\columnwidth]{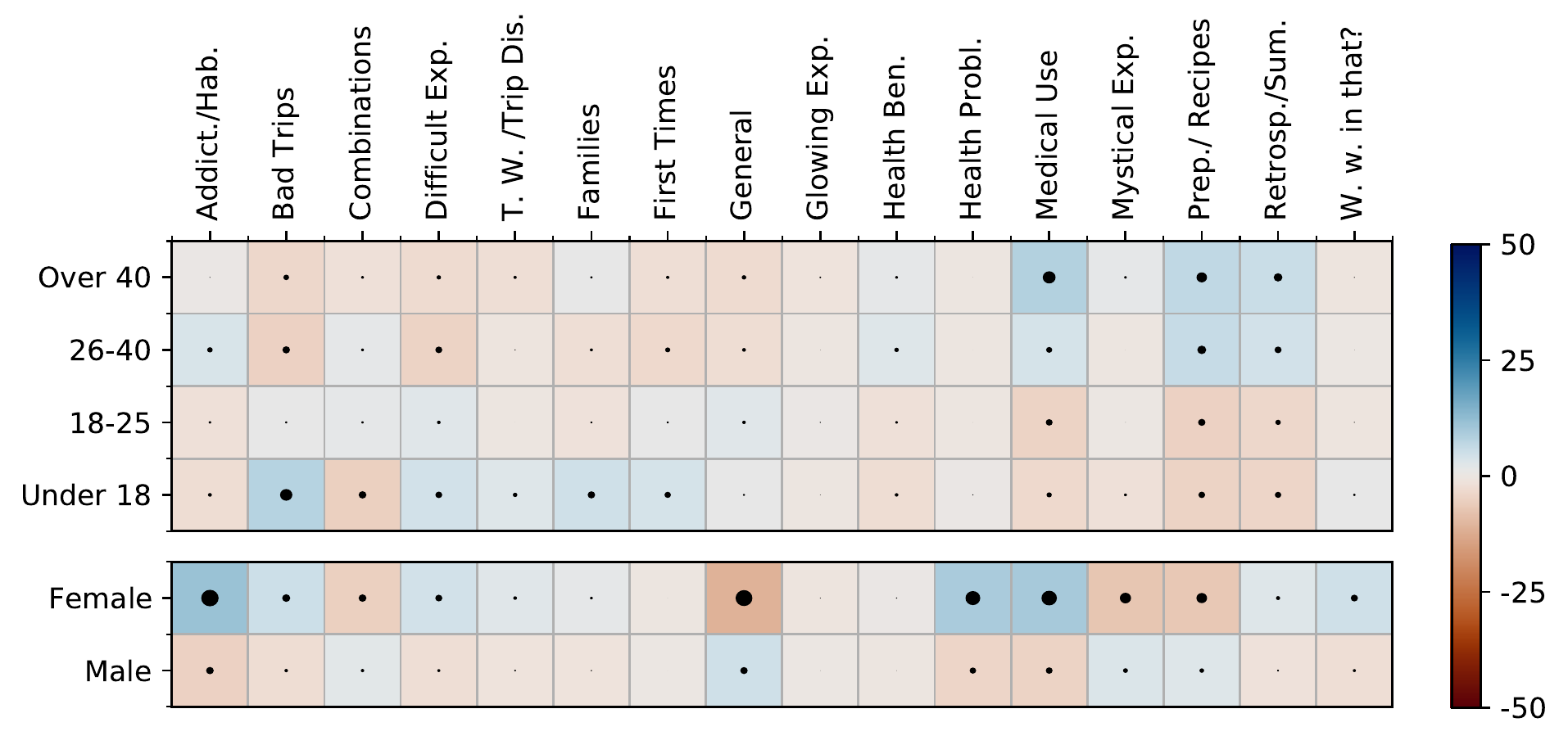}
   \caption{Chi-square distances of age group-category pairs and gender-category pairs. Legend and circle dimensions correspond to figure \ref{fig:cat_drug_residuals}.
   }
   \label{fig:cat_agesex_residuals}
\end{figure}

Figure \ref{fig:cat_agesex_residuals} shows the chi-square distance for each age-category pair. There was a significant relationship between age group and category ($\chi^2(45, N=20,758)=689, p<0.01$). The exact chi-square distances and contributions can be found in section C of the online appendix.\footnotemark[11]%\footnotemark[\ref{foot:appendix}]

First, we find that \textit{older people share more long time experiences}. It is not surprising that people above 25 have positive correlations to `Retrospective/Summary' and `Preparation/Recipes' as well as a weak positive correlation to the category `Addiction/Habituation'. Due to a higher age they had more years in which they could have experienced drug consumption; therefore, they should be able to share more drug consumption insights and to write more retrospective reports than younger people. Furthermore, they have a higher probability of having experienced an addiction and are therefore more able to report about it. 

Second, we find that \textit{older people report more about using drugs for medical purposes}. The two older groups showed positive correlations with `Medical Use', while the two younger groups showed negative correlations with it. One explanation for this is, that older people likely experience more medical issues than younger people. Another is, that they might be more motivated to contribute data about the medical use of certain substances (rather than writing entertaining stories of recreational use).

Third, we find that \textit{younger people report more about negative experiences}. The two younger groups showed positive correlations with `Bad Trips' and `Difficult Experiences', and people below 18 also had more reports in the category `Train Wrecks/Trip Disasters' than expected under an independence null hypothesis. There are several potential explanations for this: First, younger people may have less experience with taking psychoactive drugs than older drug consumers. Therefore, they probably know less about their own limits, may take higher doses than appropriate, have a lower tolerance to drugs, take these substances in less ideal settings, and are more likely to be overwhelmed by the pharmacological effects of drugs. In addition, younger people may have different motivations for (reporting) drug consumption than older people, and therefore choose a different kind of context, substance, and dosage, which may lead to a higher chance of having a negative experience. 

Furthermore, there was a significant association between gender and category ($\chi^2(15, N=59,109)= 794, p<0.01 $). We find that \textit{females report more about using drugs for medical purposes, while at the same time report more about health problems and addiction in relation to drugs}. One possible explanation for this is that females may focus on the harm reduction approach and submit reports to warn others about the addictive potential and health consequences of (medical) substances. However, the percentage of females is lower in our data than in other studies; therefore, it is also possible that only a certain kind of female drug consumer, namely women interested in health related issues, report to Erowid. Another explanation could be that males and females might differ in their motivations for drug consumption, such that women more often take drugs for medical reasons or because of drug dependence. This is in line with existing research suggesting females report using psychoactive drugs to help with anxiety or to feel better more often than males \cite{kettner_2019}. 

In summary, our data reveals gender and age play a significant role in the motivation for and interpretation of drug consumption. While younger people report more about negative drug experiences, older people and females report more about health related aspects like medical use and addiction.

\section{Predictability of Drug Experience Outcome}
Drug use is a complex interplay of pharmacological and psychological processes. The substances consumed as well as the dosage, the consumers psychological and physical status, the location and many more factors may affect the outcome of a drug experience. While consumers may take precautions to increase their probability of having a positive drug experience and avoid having a negative experience or becoming addicted, their success seems to be unpredictable.

%But is it really unpredictable? 
Surprisingly, we found regularities with all these different drug experiences: For example, LSD seems to be comparatively highly associated with negative drug experiences, while MDMA seems to be comparatively highly associated with positive ones. Furthermore, there are many variables, like dosages or settings, which we have not analyzed yet and which may reveal further patterns. Are these regularities strong enough to predict the outcome of a drug experience?

In this section we use information about the substances a user has consumed, the dosage amount, user demographics, context, and location during the experience to classify whether or not an experience will be 1) `glowing' (=positive), 2) `difficult' (=negative) and 3) related to addiction. The goal here is not `prediction' per se, but `predictability' \cite{martin2016}. In other words, we want to use predictive performance as a goodness-of-fit measure to see the maximum amount of variance an explanatory model might aspire to explain. This helps to gain deeper insights into the level of complexity drug use encompasses: If the outcomes are predictable, this would show that there are specific patterns leading to a positive, negative or addiction experience and that users therefore can influence the outcome of their drug experience by making specific choices about the substance and setting. 

\subsection{Models}
\subsubsection{Labels.}
We created three classification tasks:
For the first task, we chose the variable `Glowing Experience' (0/1) as label. We treated this label as a marker for whether a report was about a positive experience, as this category had the highest average sentiment score. For the second task, we chose `Difficult Experiences' (0/1) as label. We treated this label as a marker for whether a report was about a rather negative experience, as this was one of the four most negative categories as evaluated by the sentiment analysis and had more than twice as many reports than the other three most negative categories. For the third task, we chose `Addiction/Habituation' (0/1) as label. While this category is not the same as a clinical diagnosis of addiction, exploring this label provides valuable information about experiences that domain experts classify as being about dependency issues.

\subsubsection{Data.}
For each classification task, we excluded all reports which were not about at least one of the top ten substances or which had no consumer weight or gender information included. For the experience tasks, we excluded all reports which were about one of the top ten drugs, but had no information about the dosage of this drug. This drastically decreased the dataset. Table \ref{table:feats} shows the dataset sizes for all models. The datasets were moderately unbalanced for the experience tasks (1=19\%,0=81\%) and highly unbalanced for the addiction model (1=5\%, 0=95\%).

\begin{table*}[ht]
\footnotesize
\center
\begin{tabular}{|p{2.5cm}|p{4.5cm}|p{4.5cm}|p{4.5cm}|}
%\begin{tabular}{|c|c|c|c|}
\hline
Label & Glowing or Not & Difficult or Not & Addiction or Not \\
\hline
Dataset & 
7,414 Reports \newline (1=19\%,0=81\%) & 
7,414 Reports \newline (1=19\%, 0=81\%) &
14,113 Reports \newline (1=5\%, 0=95\%) \\ 
\hline
Features&
- Drugs \& drug combinations (142)\newline 
- Context/location (9)\newline 
- Author gender (1)\newline 
- Author weight (1)\newline 
- Drug dosages (10) & 
- Drugs \& drug combinations (142)\newline
- Context/location (9)\newline
- Author gender (1)\newline
- Author Weight (1)\newline
- Drug dosages (10) & 
- Drugs \& drug combinations (142)\newline
- Context/location (9)\newline
- Author gender (1)\newline 
- Author Weight (1) \\ 
\hline
Feat. Sel.  Model&
Lasso ($\alpha$=0.001)& 
Lasso ($\alpha$=0.991)& 
Lasso ($\alpha$=0.001)\\
\hline
\end{tabular}
\caption{Size of dataset, with balance of positive and negative reports in parentheses, features and feature selection model for the three classification tasks.}
\label{table:feats}
\end{table*}

\begin{table*}
\footnotesize
\center
\begin{tabular}{|C{1.4cm}|C{1.83cm}|C{1.83cm}|C{1.83cm}|C{1.83cm}|C{1.83cm}|C{1.83cm}|C{1.83cm}|}
\hline
Model & 
\makecell{ Majority \\ Vote} &
\makecell{ Random \\ Forest} &
\makecell{ Logistic \\ Regression} &
\makecell{ Linear Disc. \\ Analysis} &
\makecell{ K-Neighbors\\ Classifier} &
\makecell{ Decision\\ Tree} &
\makecell{ Gaussian\\ NB} \\
\hline
\makecell{Glowing \\ or Not} &
\makecell{0.0 \\ (0.81, 0.0)} &
\makecell{0.38 \\ (0.76, 0.156)} &
\makecell{0.13 \\ (0.81, 0.017)} &
\makecell{\textbf{0.31} \\ \textbf{(0.81, 0.101)}} &
\makecell{0.28 \\ (0.76, 0.087)} &
\makecell{0.38 \\ (0.74, 0.16)} &
\makecell{0.38 \\ (0.31, 0.885)} \\
\hline
\makecell{Difficult \\ or Not} &
\makecell{0.0 \\ (0.81, 0.0)} &
\makecell{\textbf{0.06} \\ \textbf{(0.81, 0.004)}} &
\makecell{0.0 \\ (0.81, 0.0)} &
\makecell{\textbf{0.06} \\ \textbf{(0.81, 0.004)}} &
\makecell{0.0 \\ (0.81, 0.0)} &
\makecell{0.0 \\ (0.81, 0.0)} &
\makecell{0.18 \\ (0.8, 0.033)} \\
\hline
\makecell{Addiction \\ or Not} &
\makecell{0.0 \\ (0.95, 0.0)} &
\makecell{0.41 \\ (0.94, 0.174)} &
\makecell{0.19 \\ (0.95, 0.038)} &
\makecell{0.7 \\ (0.93, 0.515)} &
\makecell{0.32 \\ (0.93, 0.106)} &
\makecell{\textbf{0.37} \\ \textbf{(0.95, 0.136)}} &
\makecell{0.7 \\ (0.55, 0.917)} \\
\hline
\end{tabular}
\caption{GMean for each classification task and model, with accuracy and sensitivity in parentheses. The best performing model (highest GMean with same accuracy as majority vote model) for each classification task is bolded.}
\label{table:class_results}
\end{table*}

\subsubsection{Features and Feature Selection.}
Table \ref{table:feats} shows the feature sets we used for each model. The construction of each feature is described in section \ref{section:data_clean}. As described in section \ref{section:cat_des}, reports about addiction were often written as summaries of a longer time period and therefore included no dosage information. Consequently, we did not include dosage amount as a feature for the addiction model.

From these sets of features, we chose all variables which had the strongest relationship with the target variable using Lasso regression. For each model, we split the dataset into a training set and a test set with a ratio of 4:1. We then assessed the optimal regularization parameter $\alpha$ for the lasso regression using Pythons LassoCV\footnote{\label{foot:scikit}https://scikit-learn.org/stable/} on the training set. We chose 10-fold cross validation with $\alpha$ values from 0.001 to 10 with a 0.01 step size. We then included Lasso regression with the particular alpha as a feature selection part in a scikit pipeline.\footnotemark[14] The selected alpha values are shown in table \ref{table:feats}; the resulting features for each model can be found in section D in the online appendix.\footnotemark[11]%\footnotemark[\ref{foot:appendix}]

\subsubsection{Classifiers.}
 We used six different linear and nonlinear classifiers: A random forest; logistic regression, with `liblinear' as a solver; Linear Discriminant Analysis; $k$-nearest-neighbors; scikit's decision tree, an optimized classification and regression tree (CART) algorithm; and a Gaussian Naive Bayes.\footnotemark[14] %\footnotemark[\ref{foot:scikit}] 
 Each model was included as a classifier in a scikitlearn pipeline as second element after the feature selection part, and evaluated using the standard parameters given in scikit-learn (version 0.23.2).\footnotemark[14] %\footnotemark[\ref{foot:scikit}]

\subsection{Model Evaluation}
We compared each model to the majority vote model. As the datasets were highly imbalanced, accuracy alone would be not informative for assessing the models performance. Therefore, we used the Geometric Mean (GMean), which is defined as $\sqrt{sensitvity*specitivity}$. The GMean is low, when either the prediction performance for the majority class or the prediction performance for the minority class (or both) is low. We were searching for a model, which would at least preserve the accuracy of the majority vote model, but had a better performance regarding the GMean.

\subsection{Results}
Table \ref{table:class_results} presents the GMean, as well as the accuracy and sensitivity for each classification task and model. 
We find that given our data and models \textit{the outcome of a drug experience is not predictable}. For each classification task, the best performing model reached a GMean less than 0.5. These low GMean values were caused by a low sensitivity: Only 10\% of reports, which were about a `glowing' experience, could be detect as such by our model. The same held for difficult experiences (sensitivity of 0.4\%) and addiction experiences (sensitivity of 14\%). 

The results indicate that the relationship between drug consumption and subjective outcome is highly complex and has a low degree of predictability. Hence consumers cannot simply choose a specific drug, dosage and setting, in order to ensure a positive experience and to prevent a difficult or addiction experience. These findings emphasize the danger in drug consumption: We have seen that specific drug experiences are associated with the drugs themselves, age groups and gender. Consumers can therefore assess their vulnerability and even influence their probability of having a specific experience by choosing a particular drug. But in the end, the outcome of drug consumption is still somewhat unpredictable and therefore remains a risk.

\section{Discussion and Conclusion}
The Erowid Experience Report collection gave us the unique possibility to get a consumer perspective on drug consumption. We have shown that this data can reveal valuable information about the relationship between drug consumption variables and the characteristics of a drug experience. 

Our research shed light on the subjective evaluation of drug experiences. We found that negative drug experiences are more prevalent for younger people and LSD users, while positive experiences occur more often with MDMA or first time consumption. Moreover, we gained deeper insights into health consequences of drug use: We found that reported health problems and addiction are linked to specific substances and that females report more often about these topics. Finally, our research highlighted the risk of drug consumption: Even when consumers could control the substance, dosage and situational factors, it is unpredictable, whether their experience will be joyful, difficult or associated with addiction.
 
\subsection{Limitations}
Although subjective experience reports can reveal fascinating patterns, they should be analyzed with care. First, we do not know how the sample of drug consumers, whose experiences are published on Erowid, compares to larger populations of drug consumers. As we have shown, a higher percentage of Erowid contributors are male and focus more on psychedelic drugs than drug consumers identified by national studies. They are also likely more engaged in systematic exploration with and reflection on psychoactive substances. In addition, there might be a selection bias in the data, e.g. as reviewers might favour reports about unusual substances, and/or a reporting bias, e.g. as contributors may only write about topics, substances and effects which they define as interesting.

Second, our synthesized metadata may contain errors, especially the drug dosages. Standardizing drug dosages is a challenging task which requires extensive domain knowledge. Even if users reported dosages in standardized ways, it remains unclear whether this information is correct, as users are not always aware of the specific content and composition of the substance consumed. For example, Vrolijk et. al (\citeyear{vrolijk_2017}) compared user-generated online information on ecstasy tablets to information from the validated Dutch Drugs Information and Monitoring System (DIMS) Database, and found that users tend to overestimate MDMA concentration and, in 15.3\% of cases, provided dangerously wrong information. The existence of the `What was in that?' category confirms information gaps among Erowid users as well.

Third, the prediction results are bound to the data, classifiers and parameters used. Although we tried a wide range of models, further research may find others which reveal better results and may show other factors, with which consumers could control the outcome of their drug experience. 

Fourth, it should be emphasized that Erowid has collected Experience Reports for more than 25 years and even extracted some older reports from books and journals, dating back as early as 1848. Over this time span the substance availability, drug composition and drug consumption patterns have certainly changed, which may impact our results. Further research could study temporal trends on Erowid to gain more information on the history of drug consumption.

\subsection{Implications and Future Work}
\label{section:impl}
The increasing complexity of the world's drug landscape has brought new challenges for drug consumers, medical personnel, social workers, institutions and researchers \cite{arillotta_2020,dagnone_2015,schifano_2020}. Quantitatively analyzing experience reports can help to shed light on drug-category associations, which need to be more in the focus of research. In our study for example, we found a  positive  association  between `Addiction/Habituation' and dextromethorphan (DXM), a substance which is legally available  over-the-counter in the  United States.\footnote{\label{drugchart}https://www.drugabuse.gov/drug-topics/commonly-used-drugs-charts} While the withdrawal symptoms of DXM are largely unknown,\footnotemark[15] there have been sporadic clinical reports about patients suffering from DXM dependence \cite{miller_2005,mutschler_2010}. Further research could investigate more deeply the relationships we have found, for example by manually inspecting the reports and conducting psychopharmacological studies. Moreover, the chi-square approach we used in this paper can be applied to other substances and topics, and help to generate new hypotheses for research. 

In addition, studying Erowid Experience Reports can help to obtain a deeper sociological understanding of drug use. Our results indicate age and gender play a significant role in the motivation for and interpretation of drug consumption. This should allow further research to investigate whether there are also demographic differences in the choices drug consumers make, for example regarding the strength of drug dosages or the setting for drug consumption.

Finally, Erowid's labeling and categorization process allows the analysis of a great variety of topics, which were beyond the scope of this paper. 
Many of the categories described, such as `Families' or `Mystical Experiences', show a great potential for further research, as they can provide new insights into how people experience drug consumption.

\fontsize{9.0pt}{10.0pt} \selectfont
\bibliography{ICWSM_2021_Erowid.bib}

\begin{thebibliography}{42}
\providecommand{\natexlab}[1]{#1}
\providecommand{\url}[1]{\texttt{#1}}
\providecommand{\urlprefix}{URL }
\expandafter\ifx\csname urlstyle\endcsname\relax
  \providecommand{\doi}[1]{doi:\discretionary{}{}{}#1}\else
  \providecommand{\doi}{doi:\discretionary{}{}{}\begingroup
  \urlstyle{rm}\Url}\fi

\bibitem[{Andersson and Kjellgren(2017)}]{andersson_slippery_2017}
Andersson, M.; and Kjellgren, A. 2017.
\newblock The slippery slope of flubromazolam: Experiences of a novel
  psychoactive benzodiazepine as discussed on a Swedish online forum.
\newblock \emph{Nordic Studies on Alcohol and Drugs} 34(3): 217--229.

\bibitem[{Arillotta et~al.(2020)Arillotta, Schifano, Napoletano, Zangani,
  Gilgar, Guirguis, Corkery, Aguglia, and Vento}]{arillotta_2020}
Arillotta, D.; Schifano, F.; Napoletano, F.; Zangani, C.; Gilgar, L.; Guirguis,
  A.; Corkery, J.~M.; Aguglia, E.; and Vento, A. 2020.
\newblock Novel opioids: Systematic web crawling within the e-psychonauts'
  scenario.
\newblock \emph{Frontiers in Neuroscience} 14: 149.

\bibitem[{Baggott et~al.(2011)Baggott, Coyle, Erowid, Erowid, and
  Robertson}]{baggott_2011}
Baggott, M.; Coyle, J.; Erowid, E.; Erowid, F.; and Robertson, L. 2011.
\newblock Abnormal visual experiences in individuals with histories of
  hallucinogen use: A web-based questionnaire.
\newblock \emph{Drug and Alcohol Dependence} 114(1): 61--67.

\bibitem[{Baggott et~al.(2010)Baggott, Erowid, Erowid, Galloway, and
  Mendelson}]{baggott_2010}
Baggott, M.~J.; Erowid, E.; Erowid, F.; Galloway, G.~P.; and Mendelson, J.
  2010.
\newblock Use patterns and self-reported effects of Salvia divinorum: An
  internet-based survey.
\newblock \emph{Drug and Alcohol Dependence} 111(3): 250--256.

\bibitem[{Berning and Hardon(2016)}]{berning_2016}
Berning, M.; and Hardon, A. 2016.
\newblock Educated guesses and other ways to address the pharmacological
  uncertainty of designer drugs: An exploratory study of experimentation
  through an online drug forum.
\newblock \emph{Contemporary Drug Problems} 43(3): 277--292.

\bibitem[{Bilgrei(2019)}]{bilgrei_2019}
Bilgrei, O.~R. 2019.
\newblock Community‐consumerism: Negotiating risk in online drug communities.
\newblock \emph{Sociology of Health \& Illness} 41(5): 852--866.

\bibitem[{Bonson, Buckholtz, and Murphy(1996)}]{bonson_1996}
Bonson, K.~R.; Buckholtz, J.~W.; and Murphy, D.~L. 1996.
\newblock Chronic administration of serotonergic antidepressants attenuates the
  subjective effects of {LSD} in Humans.
\newblock \emph{Neuropsychopharmacology} 14(6): 425--436.

\bibitem[{Chary, Yi, and Manini(2018)}]{chary_2018}
Chary, M.; Yi, D.; and Manini, A.~F. 2018.
\newblock Candyflipping and other combinations: Identifying drug-drug
  combinations from an online forum.
\newblock \emph{Frontiers in Psychiatry} 9: 135.

\bibitem[{D'Agnone(2015)}]{dagnone_2015}
D'Agnone, O. 2015.
\newblock What have we learned and what can we do about NPS?
\newblock \emph{Drugs and Alcohol Today} 15(1): 28--37.

\bibitem[{Duxbury(2018)}]{duxbury_2018}
Duxbury, S.~W. 2018.
\newblock Information creation on online drug forums: How drug use becomes
  moral on the margins of science.
\newblock \emph{Current Sociology} 66(3): 431--448.

\bibitem[{Erowid and Erowid(2006)}]{erowid_10}
Erowid, E.; and Erowid, F. 2006.
\newblock The Value of Experience: Erowid's Collection of First-Person
  Psychoactive Reports.
\newblock \emph{Erowid Extracts} 10: 14--19.

\bibitem[{Erowid, Thyssen, and Erowid(2018)}]{erowid_30}
Erowid, F.; Thyssen, S.; and Erowid, E. 2018.
\newblock Mountains of Experience: The process of reviewing 112,000
  first-person reports about the use of psychoactives.
\newblock \emph{Erowid Extracts} 30: 6--7.

\bibitem[{{European Monitoring Centre for Drugs and Drug
  Addiction}(2020)}]{eu_drug_report_2020}
{European Monitoring Centre for Drugs and Drug Addiction}. 2020.
\newblock \emph{European Drug Report 2020: Trends and Developments}.
\newblock Publications Office of the European Union, Luxembourg.

\bibitem[{Gamma et~al.(2005)Gamma, Jerome, Liechti, and Sumnall}]{gamma_2005}
Gamma, A.; Jerome, L.; Liechti, M.~E.; and Sumnall, H.~R. 2005.
\newblock Is ecstasy perceived to be safe? {A} critical survey.
\newblock \emph{Drug and Alcohol Dependence} 77(2): 185--193.

\bibitem[{Hutto and Gilbert(2014)}]{Revision_Hutto2014}
Hutto, C.; and Gilbert, E. 2014.
\newblock VADER: A Parsimonious Rule-based Model for Sentiment Analysis of
  Social Media Text.
\newblock \emph{Proceedings of the 8th International Conference on Weblogs and
  Social Media, ICWSM 2014} .

\bibitem[{Jarnow(2016)}]{jarnow_2016}
Jarnow, J. 2016.
\newblock Why Psychedelic History Matters.
\newblock \emph{VolteFace} .

\bibitem[{Karila et~al.(2016)Karila, Billieux, Benyamina, Lancon, and
  Cottencin}]{karila_2016}
Karila, L.; Billieux, J.; Benyamina, A.; Lancon, C.; and Cottencin, O. 2016.
\newblock The effects and risks associated to mephedrone and methylone in
  humans: A review of the preliminary evidences.
\newblock \emph{Brain Research Bulletin} 126, Part 1: 61--67.

\bibitem[{Kettner, Mason, and Kuypers(2019)}]{kettner_2019}
Kettner, H.; Mason, N.~L.; and Kuypers, K. P.~C. 2019.
\newblock Motives for classical and novel psychoactive substances use in
  psychedelic polydrug Users.
\newblock \emph{Contemporary Drug Problems} 46(3): 304--320.

\bibitem[{Krieg et~al.(2016)Krieg, Berning, Colombo, Azzi, and
  Hardon}]{krieg_2016}
Krieg, L.~J.; Berning, M.; Colombo, G.; Azzi, M.; and Hardon, A. 2016.
\newblock Visualising {Erowid}: A data driven anthropological research on
  drugs.
\newblock Chemical Youth, University of Amsterdam.

\bibitem[{Krieg, Berning, and Hardon(2017)}]{krieg_2017}
Krieg, L.~J.; Berning, M.; and Hardon, A. 2017.
\newblock Anthropology with algorithms?
\newblock \emph{Medicine Anthropology Theory} 4(3).

\bibitem[{Marlatt(1996)}]{marlatt1996}
Marlatt, G.~A. 1996.
\newblock Harm reduction: Come as you are.
\newblock \emph{Addictive Behaviors} 21(6): 779--788.

\bibitem[{Martin et~al.(2016)Martin, Hofman, Sharma, Anderson, and
  Watts}]{martin2016}
Martin, T.; Hofman, J.~M.; Sharma, A.; Anderson, A.; and Watts, D.~J. 2016.
\newblock Exploring limits to prediction in complex social systems.
\newblock In \emph{Proceedings of the 25th International Conference on World
  Wide Web}, 683--694.

\bibitem[{Miller(2005)}]{miller_2005}
Miller, S. 2005.
\newblock Dextromethorphan psychosis, dependence and physical withdrawal.
\newblock \emph{Addiction Biology} 10(4): 325--327.

\bibitem[{Mutschler et~al.(2010)Mutschler, Koopmann, Grosshans, Hermann, Mann,
  and Kiefer}]{mutschler_2010}
Mutschler, J.; Koopmann, A.; Grosshans, M.; Hermann, D.; Mann, K.; and Kiefer,
  F. 2010.
\newblock Dextromethorphan withdrawal and dependence syndrome.
\newblock \emph{Deutsches Arzteblatt international} 107(30): 537–540.

\bibitem[{Newman et~al.(2016)Newman, Denton, Walker, and Grewal}]{newman_2016}
Newman, M.; Denton, G.; Walker, T.; and Grewal, J. 2016.
\newblock The experience of using synthetic cannabinoids: A qualitative
  analysis of online user self-reports.
\newblock \emph{European Psychiatry} 33: S309--S310.

\bibitem[{Norman, Grace, and Lloyd(2014)}]{norman_2014}
Norman, J.; Grace, S.; and Lloyd, C. 2014.
\newblock Legal high groups on the internet --- The creation of new organized
  deviant groups?
\newblock \emph{Drugs: Education, Prevention and Policy} 21(1): 14--23.

\bibitem[{Pal et~al.(2013)Pal, Balt, Erowid, Erowid, Baggott, Mendelson, and
  Galloway}]{pal_2013}
Pal, R.; Balt, S.; Erowid, E.; Erowid, F.; Baggott, M.; Mendelson, J.; and
  Galloway, G. 2013.
\newblock Ketamine is associated with lower urinary tract signs and symptoms.
\newblock \emph{Drug and Alcohol Dependence} 132(1--2): 189--194.

\bibitem[{Paul et~al.(2016)Paul, Chisolm, Johnson, Vandrey, and
  Dredze}]{paul_2016}
Paul, M.~J.; Chisolm, M.~S.; Johnson, M.~W.; Vandrey, R.~G.; and Dredze, M.
  2016.
\newblock Assessing the validity of online drug forums as a source for
  estimating demographic and temporal trends in drug use.
\newblock \emph{Journal of Addiction Medicine} 10(5): 324--330.

\bibitem[{Rhumorbarbe et~al.(2019)Rhumorbarbe, Morelato, Staehli, Roux,
  Jaquet-Chiffelle, Rossy, and Esseiva}]{rhumorbarbe_2019}
Rhumorbarbe, D.; Morelato, M.; Staehli, L.; Roux, C.; Jaquet-Chiffelle, D.-O.;
  Rossy, Q.; and Esseiva, P. 2019.
\newblock Monitoring new psychoactive substances: Exploring the contribution of
  an online discussion forum.
\newblock \emph{Internat. Journal of Drug Policy} 73: 273--280.

\bibitem[{Rolando and Beccaria(2019)}]{rolando_2019}
Rolando, S.; and Beccaria, F. 2019.
\newblock ``The junkie abuses, the psychonaut learns'': A qualitative analysis
  of an online drug forum community.
\newblock \emph{Drugs and Alcohol Today} 19(4): 282--294.

\bibitem[{{RTI International}(2019)}]{nsdhu_codebook_2018}
{RTI International}. 2019.
\newblock \emph{2018 National Survey on Drug Use and Health Public Use File
  Codebook}.
\newblock Center for Behavioral Health Statistics and Quality, Substance Abuse
  and Mental Health Services Administration.

\bibitem[{Ruths and Pfeffer(2014)}]{Ruths2014}
Ruths, D.; and Pfeffer, J. 2014.
\newblock Social Media for Large Studies of Behavior.
\newblock \emph{Science} 346(6213): 1063--1064.

\bibitem[{Schifano(2020)}]{schifano_2020}
Schifano, F. 2020.
\newblock Analyzing the open/deep web to better understand the new/novel
  psychoactive substances.
\newblock \emph{Brain Sciences} 10(3): 146.

\bibitem[{Soussan and Kjellgren(2014)}]{soussan_2014}
Soussan, C.; and Kjellgren, A. 2014.
\newblock Harm reduction and knowledge exchange---a qualitative analysis of
  drug-related Internet discussion forums.
\newblock \emph{Harm Reduction Journal} 11(1): 25.

\bibitem[{Stanciu, Penders, and Rouse(2016)}]{stanciu2016}
Stanciu, C.~N.; Penders, T.~M.; and Rouse, E.~M. 2016.
\newblock Recreational use of dextromethorphan, ``Robotripping''---A brief
  review.
\newblock \emph{The American Journal on Addictions} 25(5): 374--377.

\bibitem[{Stone and Shirley-Beavan(2018)}]{stone2018}
Stone, K.; and Shirley-Beavan, S. 2018.
\newblock The Global State of Harm Reduction 2018.
\newblock Technical report, Harm Reduction International, London.

\bibitem[{{Substance Abuse and Mental Health Services
  Administration}(2019)}]{nsdhu_2018}
{Substance Abuse and Mental Health Services Administration}. 2019.
\newblock \emph{Key Substance Use and Mental Health Indicators in the United
  States}.
\newblock HS Publ. No. PEP19-5068, NSDUH Series H-54.

\bibitem[{Swogger et~al.(2015)Swogger, Hart, Erowid, Erowid, Trabold, Yee,
  Parkhurst, Priddy, and Walsh}]{swogger_2015}
Swogger, M.~T.; Hart, E.; Erowid, F.; Erowid, E.; Trabold, N.; Yee, K.;
  Parkhurst, K.~A.; Priddy, B.~M.; and Walsh, Z. 2015.
\newblock Experiences of Kratom Users: A Qualitative Analysis.
\newblock \emph{Journal of Psychoactive Drugs} 47(5): 360--367.

\bibitem[{{United Nations}(2018)}]{world_drug_report_2018}
{United Nations}. 2018.
\newblock \emph{World Drug Report 2018}.
\newblock United Nations publication, Sales No. E.18.XI.9.

\bibitem[{{United Nations}(2020)}]{world_drug_report_2020}
{United Nations}. 2020.
\newblock \emph{World Drug Report 2020}.
\newblock United Nations publication, Sales No. E.20.XI.6.

\bibitem[{Vrolijk et~al.(2017)Vrolijk, Brunt, Vreeker, and
  Niesink}]{vrolijk_2017}
Vrolijk, R.~Q.; Brunt, T.~M.; Vreeker, A.; and Niesink, R. J.~M. 2017.
\newblock Is online information on ecstasy tablet content safe?
\newblock \emph{Addiction} 112(1): 94--100.

\bibitem[{Witt(2015)}]{witt_2015}
Witt, E. 2015.
\newblock The trip planners: The unusual couple behind an online encyclopedia
  of psychoactive substances.
\newblock \emph{The New Yorker,} November 16, 2015.

\end{thebibliography}

%\section{Appendix}

\end{document}